# Search for New Superconductors: An Electro-Magnetic Phase Transition in an Iron Meteorite Inclusion at 117 K

S. Guénon[1,2]*, J. G. Ramírez[1,3], Ali C. Basaran[1], J. Wampler[1], M. Thiemens[4], and Ivan K. Schuller[1]

**Affiliations:**

[1]Department of Physics and Center for Advanced Nanoscience, University of California San Diego, La Jolla CA 92093, USA.

[2]CQ Center for Collective Quantum Phenomena and their Applications in LISA+, Physikalisches Institut, Eberhard Karls Universität Tübingen, Auf der Morgenstelle 14, D-72076, Tübingen, Germany.

[3]Department of Physics, Universidad de los Andes, Bogotá 111711, Colombia

[4]Department of Chemistry and Biochemistry, University of California San Diego, La Jolla CA 92093, USA.

*Correspondence to: stefan.guenon@uni-tuebingen.de

## Abstract:

The discovery of superconductivity in pnictides and iron chalcogenides inspires the search for new iron based superconducting phases. Iron-rich meteorites present a unique opportunity for this search, because they contain a broad range of compounds produced under extreme growth conditions. We investigated a natural iron sulfide based materials (Troilite) inclusion with its associated minerals in the iron meteorite Tlacotepec. Tlacotepec formed in an asteroidal core under high pressure and at high temperature over millions of years, while insoluble sulfur rich materials segregated into inclusions during cooling along with included minerals. The search for superconductivity in these heterogeneous materials requires a technique capable of detecting minute amounts of a superconducting phase embedded in a non-superconducting matrix. We used Magnetic

Field Modulated Microwave Spectroscopy (MFMMS), a very sensitive, selective, and non-destructive technique, to search for superconductivity in heterogeneous systems.

Here, we report the observation of an electro-magnetic phase transition at 117 K that causes a MFMMS-response typical of a superconductor. A pronounced and reproducible peak together with isothermal magnetic field sweeps prove the appearance of a new electromagnetic phase below 117 K. This is very similar to the characteristic response due to flux trapping in a granular superconductor with a short coherence length. Although the compound responsible for the peak in the MFMMS-spectra was not identified, it is possibly an iron sulfide based phase, or another material heterogeneously distributed over the inclusion.

## Main Text:

The discovery of iron-based superconductors, a new class of superconducting compounds with an unconventional pairing mechanism, has rekindled the interest in the search for new superconducting materials [1-3]. Superconducting iron chalcogenides like undoped β-FeSe [4], Fe(Se$_{1-x}$Te$_x$) [5,6] or S-substituted FeTe [7] are of particular interest. The superconducting transition temperature $T_c$ of FeSe can be significantly increased by applying high external pressure [8]. Several attempts were made to increase the transition temperature either by high pressure, high temperature synthesis, or by substituting selenium with sulfur that has a smaller atomic radius [9,10]. However, in this study we search for superconductivity in naturally occurring materials instead of synthesizing new compounds. This approach has two advantages: First, we can investigate materials that have crystallized under very extreme conditions like very long cooling time, high pressure, unusual chemical compositions, and high temperature, which are difficult or impossible to obtain in a laboratory. Second, due to the intrinsic inhomogeneity of these compounds, it is possible to investigate a large variety of material phases without the need to synthesize them.

Iron meteorites are crystallized at the core of protoplanetary bodies under extreme pressure and temperature [11-16]. They predominately contain Iron and Nickel (15.9%) and Platinum (0.05%) [13] but also a large array of minerals, such as pnictides. The Tlacotepec meteorite, which this study focuses on, is an iron IVB meteorite. It has an inferred kamacite nucleation temperature of 4750 ºC and cooled at a rate of 500 ºC per million years at a pressure below 1 Gpa [13] with an internal temperature of more than 1487 ºC, which provided fast planetary differentiation time scales. The presence of excess $^{107}$Ag from extinct $^{107}$Pd ($t_{1/2}$ = 6.5 x $10^6$) during its solidification requires a time scale for formation and differentiation of less than $10^7$ years [17-19]. During formation sulfur-rich liquid is continuously segregated under high pressure into droplets forming inclusions. Troilite is the most common inclusion in the Tlacotepec meteorite with some inclusion sizes in the cm-range. The meteorite also contains chromite, sphalerite, phosphate bearing sulfides, silica, CrN, rare earth elements and native copper [20]. These conditions offer a unique opportunity to search for the presence of superconductivity in unconventional, inhomogeneous natural systems, which does not exist in terrestrial materials. However, the inhomogeneity of these materials requires a sensitive and selective technique, to detect the presence of a superconducting phase. See Supplement for greater detail on the material properties and formation of iron meteorites and Tlacotepec.

In general, superconductors exhibit an electro-magnetic phase transition (EMPT), in which two main characteristic properties change drastically when they are cooled below the critical temperature $T_c$: zero electrical resistance [21] and perfect diamagnetism [22]. The Magnetic Field Modulated Microwave Spectroscopy (MFMMS) used in this study probes for a simultaneous change in both electrical and magnetic properties of the materials under investigation. MFMMS (also referred to as magnetic field modulated microwave absorption or differential/field-dependent microwave absorption) is an established technique [23-25] although it is not as common as the standard techniques for detecting superconductivity. Because it is contactless, it is very useful for investigating fragile [26] or air sensitive samples [27]. In a series of extensive studies we have shown earlier that MFMMS has the unprecedented sensitivity of $10^{-12}$ cc of a superconducting material embedded in a non-superconducting matrix [28,29]. For these reasons, we chose MFMMS to search for traces of superconductivity in naturally occurring inhomogeneous materials, such as meteorites and minerals.

Here, we report on an electro-magnetic phase transition (EMPT) at 117 K in an iron sulfide mineral extracted from the core of an iron meteorite, which causes a MFMMS response that is typical of a superconductor.

MFMMS is based on a modified electron paramagnetic resonance spectrometer [28,29]. The material under investigation is placed in a quartz tube mounted in a continuous flow Helium cryostat, which allows for changing the temperature between 3.5 and 300 K. The sample is positioned in a 9.4 GHz microwave cavity at the maximum (minimum) microwave magnetic (electric) field. A DC magnetic field (0 Oe - 9000 Oe) is applied with an electromagnet. A pair of Helmholtz-coils allows applying a collinear AC field between -100 Oe and 100 Oe with a 100 kHz modulation frequency. For a typical MFMMS temperature scan, a small (15Oe) DC-field is applied together with a 15 Oe AC field modulated at 100 kHz, while the sample temperature is continuously ramped between two temperatures (across the expected $T_c$). The change of in reflected microwave power from the cavity due to the AC magnetic field is read out via a lock-in technique.

The MFMMS-response depends on the materials under investigation [28]. In particular, in a DC magnetic field, as the temperature is swept through the Tc, the AC magnetic field forces the material periodically in and out of the superconducting state. This "derivative" type measurement gives rise to the characteristic MFMMS peak and the high sensitivity. This characteristic behavior was observed in many superconductors independently of the type (elemental, A15, $MgB_2$, cuprates, etc).

The MFMMS was designed to suppress the response of non-superconducting materials, thus the spectra of the vast majority of non-superconducting materials are featureless. However, in some rare cases (i.e. Mn doped GaAs), a peak in the MFMMS-response is obtained due to magneto-resistive effects or, due to a magnetic field induced change in the skin depth [28]. To discriminate superconducting materials from these rare cases, complementary criteria are used:

(i) the response of all superconducting materials has the same sign. If a material causes a dip instead of a peak with respect to a reference superconductor, then superconductivity can be ruled out.

(ii) small applied magnetic fields compared to the upper critical field $H_{c2}$ should have little effect on the MFMMS peak onset associated with the critical temperature of a superconductor.

(iii) the chirality of isothermal hysteretic field scan loops (FSLs) is clockwise for all known superconductors. In contrast, hysteretic FSLs for non-SCs are generally counter-clockwise, although there are a small number of exceptions (see supplement).

Refer to [29] for a short introduction and to [28] for an extensive discussion of the MFMMS technique.

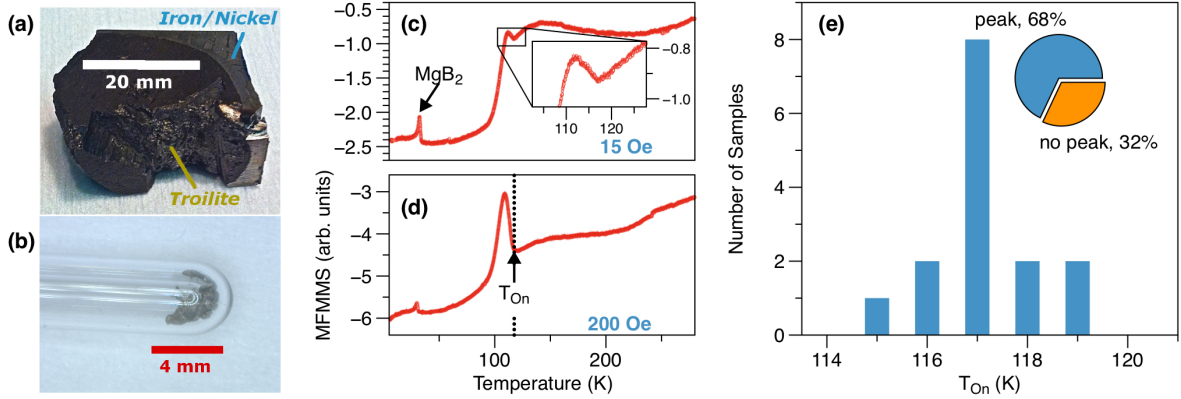

FIG.1. (color online, double column) Magnetic Field Modulated Microwave Spectroscopy (MFMMS) temperature scans of Troilite. (a) Photograph of the Tlacotepec Troilite inclusion under investigation. (b) Photograph of the sample Tla-1. (c, d) MFMMS of Tla-1 with different DC off-set fields (blue labels). The onset temperature ($T_{On}$) is indicated by a vertical dotted line. (e) Distribution of $T_{On}$ of different sample tubes. In total 22 sample tubes were investigated and 15 had a peak in vicinity of 117 K in MFMMS with 15 Oe applied DC-field.

Figure 1(a) shows a photograph of the sample under investigation. It is approximately one quarter of a Troilite inclusion from the iron meteorite Tlacotepec. We quarried out 0.5 cc of materials from the front side of the Troilite inclusion, crushed it in a mortar, and separated it into different samples for MFMMS-analysis. In Fig. 1(b) a photograph of a sample tube (Tla-1) is shown.

Figure 1(c) shows a typical MFMMS-spectrum of the material under investigation. The overall shape is a flat negative response from room temperature to 145 K followed by a rounded, s-shaped step going down, ending at 60 K. Most important, there is a peak with an onset temperature of 117 K. This peak has the same sign as the peak of a reference superconductor $MgB_2$ that was added to the samples. At a higher magnetic DC offset field the peak at 117 K is more pronounced [Fig. 1(d)]. We have investigated 22 different sample in total, 15 of which showed a peak with an onset temperature around 117 K in

MFMMS with a DC-field of 15 Oe [Fig. 1(e)]. Hence, the repeatable and reproducible peak in the MFMMS spectrum at 117 K is indicative of an EMPT satisfying criterion (i).

The shape of the background resembles the MFMMS-spectra of $Fe_3O_4$ (magnetite) found in both synthetic powders and in micrometeorites [29]. This background has not been detected in every extraterrestrial material. For instance, the MFMMS spectra that we have acquired from the Allende and Murchison meteorites, lunar and Martian rocks show essentially a flat background. In the MFMMS spectra of the Tlacotepec meteorite, the background is probably caused by traces of $Fe_3O_4$ [29].

To investigate further the origin of the signal, we subdivided the sample repeatedly and found that similar peaks appeared in the subdivisions, at the same temperature but of lower magnitude. This implies that the EMPT phase occurs within only a small fraction of the sample, distributed throughout the inclusion.

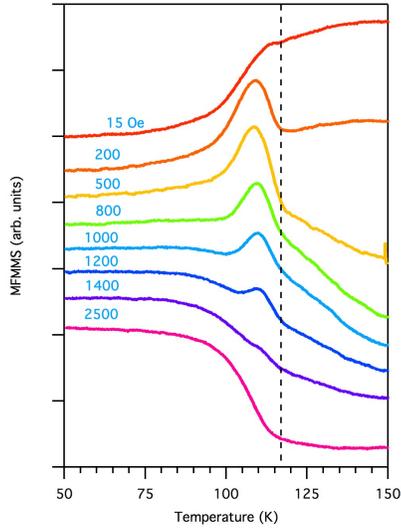

FIG 2. (color online) Magnetic Field Modulated Microwave Spectroscopy (MFMMS) data of Tla-1, with different magnetic DC off-set fields (blue labels). The spectra are shifted vertically for clarity. The vertical dashed line at 117 K is a guide to the eye that indicates the onset temperature.

A series of MFMMS-spectra from the sample Tla-1, with increasing magnetic DC offset field are shown in Fig. 2. The peak onset temperatures remain constant for different DC offset fields, although the peak heights and the background drastically change. Further, the peak is largely suppressed above 1400 Oe applied field at which the background signal becomes dominant. The fact that the peak response is independent from the background is strong evidence that the peak and the background have different origins.

Because the MFMMS of a superconductor is related to the change of the penetrating magnetic flux induced by a small magnetic AC-field, it can be suppressed by a relatively small applied magnetic field (see chapter 4 in Ramírez et al. [28]). Note that the suppression of the peak is not due to the suppression of superconductivity, instead it is due to the fact that the relative change in the penetrating magnetic flux induced by the small AC-field used (15 Oe) can no longer be detected by the MFMMS-system. Therefore, the data in Fig. 2 suggest that the criterion (ii) also holds.

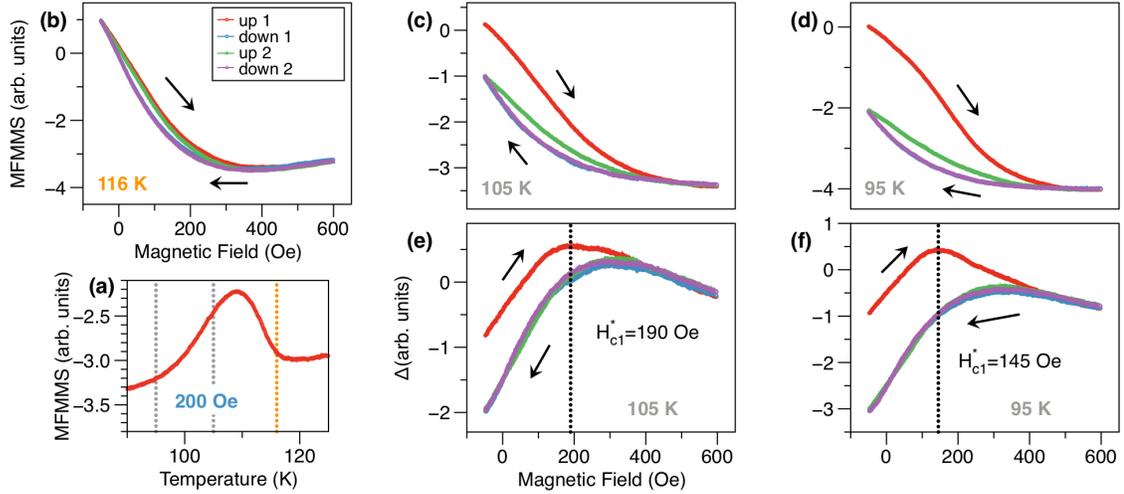

FIG 3. (color online, double column) Field scan study of Tla-1. (a) MFMMS-spectrum, blue label indicates DC off-set field, vertical lines indicate set-temperatures of field scan loops (FSLs). (b-d) Two consecutive FSLs acquired at set-temperatures indicated by labels. (e-f) Background removed FSLs (BRFSLs): FSLs at 116 K (b) subtracted from FSLs at 105 K and 95 K, respectively. The dotted vertical lines indicate the maxima of the first up sweep curves ($H_{c1}^*$).

To investigate the chirality (as mentioned above), we have acquired hysteretic FSLs following a procedure with field cooling and asymmetric field scans similar to the ones used in cuprates and spin-glass studies [33-35]. The sample was cooled from 150 K to the intended set-temperature in a low 10 Oe DC-field. After 30 minutes wait for temperature stabilization, the field was set to -50 Oe and then two consecutive FSLs between -50 Oe and 600 Oe were measured. Finally, the sample was heated to 150 K, which allowed the low field cool to the next set temperature.

Figure 3 summarizes the results from the FSLs method in the Tlacotepec Tla-1 sample. Figure 3(a) shows the MFMMS-spectrum with a pronounced peak indicating an EMPT. We acquired FSLs at different temperatures above and below the peak. FSLs typical for a set temperature above the EMPT are shown in Fig. 3(b): they have a small hysteresis, i.e. the up-sweep curve lies above the down-sweep curve, and the hysteresis of the second loop is slightly reduced. Both down-sweep curves coincide. Figure 3(c,d) show FSLs acquired at temperatures below the peak: the first up-sweep curve starts at larger values than the other curves, i.e. the first FSL is open, while the consecutive FSLs are closed. Hence, there is a training effect in the FSLs associated with the EMPT indicated by the peak in the MFMMS-spectra. In order to clearly associate the EMPT with the unknown phase the background signal above the transition was subtracted as shown in Fig. 3(e,f). This is justified by the results showed in Fig. 2, which demonstrates that the peak and the background are of different origin.

These background removed FSLs (BRFSLs) [Fig. 3(e,f)] are very similar to the spin-glass behavior observed in granular high $T_c$ superconductors and magnetic systems [33-35]. Starting at negative field values, they increase to a maximum and develop a small, negative slope for higher positive fields. More importantly, the maximum of the up-sweep curve of the first BRFSL is at a lower field value than the maxima of the down sweep curve and the other consecutive curves. Sastry et al. [34], compared the FSL characteristics of spin-glasses and vortex-glasses of granular high $T_c$ superconductors. They found very similar FSLs in both cases with one clear distinction: for a superconductor, the maximum of the first up-sweep curve is at lower field values than the maximum of the down-sweep curve, while it is the opposite for a spin-glass. For superconductors, the difference in the field positions of the up and down sweep maxima was associated with flux trapping [31]. During the first up-sweep after cool-down, magnetic flux is trapped in the superconductor and the maxima of the consecutive field sweeps are shifted to higher field values. For spin-glasses, the first up-sweep changes the spin configuration and the consecutive down-sweep curve is shifted to lower field values, as it is the case for every ferromagnetic system.

According to the Josephson Junction network model for superconducting clusters [33,36], the position of the maximum in the first up-sweep curve is related to $H_{c1}^*$ - the magnetic field at which magnetic flux quanta start to enter the loops of the Josephson Junction network: $S=\phi_0/2\ H_{c1}^*$, where S is the average surface area of the loops and $\phi_0$ the magnetic flux quantum. In our study the magnetic fields corresponding to the maxima in the first up-sweep curves are more than one order of magnitude larger than in Blazey et al. [33]. However, these fields correspond to loop areas of 0.04-0.07 $\mu m^2$ comparable to the 0.1 $\mu m^2$, which were determined by measuring Shapiro steps at YBCO samples [37], and could be realistic.

The BRFSLs in Fig. 3(e,f) have a clockwise chirality and satisfy criteria (iii). Although the chirality criteria is not universal (as discussed in the supplemental material), it is unlikely that a non-superconducting material shows all the characteristics of a granular high $T_c$ superconductor in such a complex cooling and field sweep sequence.

Several caveats should be emphasized regarding the possibility of this being evidence for superconductivity in a meteorite.

i) So far we have only acquired MFMMS-temperature scans, in which the peak was on a negative background, and the background varied from sample to sample. It can be concluded that the materials under investigation consist of a variety of material phases with different microwave responses. We cannot exclude that the background deforms the peak considerably. However, it is safe to assume that the peak is associated with an EMPT for following reasons: First, the peak is reproducible for different samples with a sharp onset. Second, in the temperature scan series with varying magnetic DC-fields (shown in figure 2), the onset-position is constant, while the peak height varies continuously, and third the training effect indicated by the gap in the FSLs (shown in figure 3) is associated with the peak in the temperature scans.

ii) Beside the microwave study discussed, we have tested these materials with additional methods like SQUID and AC-susceptibility (see supplement). We found no evidence for an EMPT in vicinity of 117 K with these methods. However, using MFMMS we have found compelling evidence for the existence of an EMPT at this temperature. The lack of signature of the EMPT in some measurements is similar to earlier studies of very small superconducting fraction lithographically prepared [28] and in small quantities of superconducting powders mixed into mostly magnetic samples [29]. In both of these cases, the sensitivity of the MFMMS was much higher than in SQUID magnetometry.

iii) Since the chemistry of meteorites is very complex, it is possible that they contain a superposition of several unidentified, unknown phases, which mimic the response of a superconductor. We have performed extensive exploration of many different materials system without such an observation, so we consider this a very unlikely scenario.

We are working on a divide and conquer methodology, in which the sample is subdivided multiple times until the response becomes unique. These measurements, although quite tedious, presumably will permit to isolate the material phase causing the peak. Then it will be possible to identify the responsible compound and to test for zero resistance and the Meissner effect.

Conclusion:

We found an EMPT with a transition temperature $T_c$ of 117 K in materials extracted from an inclusion of the Tlacotepec meteorite. The response of the EMPT in MFMMS meets all criteria of a superconducting response. In particular, the occurrence of the peak in the temperature dependence of the MFMMS signal, and the shape and the training effect of the BRFSLs are signatures of a superconducting transition. Preliminary energy dispersive X-ray spectroscopy measurements (see supplemental material) indicate that the materials under investigation consist predominantly of iron and sulfur, though also a wide variety of other trace minerals. So far, we could not identify the material phase responsible for the EMPT. It possibly is a very small iron sulfide based phase unevenly distributed throughout the inclusion, though numerous other mineralogical species may be operative. It could be that this phase only forms under the extreme growth conditions found in iron meteorites (high temperature, high pressure, very long cooling period) and that it contains additional trace elements. Considering the confirmed superconductivity of the iron chalcogenides FeTe and FeSe, it would not be surprising if this iron sulfide based phase is superconducting. To ultimately prove superconductivity, it is necessary to identify the material phase associated with the EMPT, to confirm zero resistance and the Meissner effect with complementary techniques.

**Acknowledgments:**

**Research supported by an AFOSR grant FA9550-14-1-0202. We thank Neil Dilley from Quantum Design for helping us with the SQUID and AC-susceptibility measurements. S.G. thanks József Fortágh for giving him the opportunity to finish this work. We thank Harold Weinstock for his original idea on the search for superconductivity in unconventional materials.**

**Contributions:**
This is a highly collaborative research. I.K.S. generated the idea to develop the MFMMS for the systematic search for superconductivity. The equipment was setup and tested by I.K.S., A.C.B. and J.G.R.. I.K.S. started the collaboration with M.T. who provided the samples originally obtained from the Field Museum of Natural History. M.T. suggested the Tlacotepec materials to search for superconductivity.
S.G. generated the idea to search for superconductors in extraterrestrial materials and made most of the measurements and data analysis. In particular, S.G. discovered the similarities with the field sweep studies of spin glasses and granular superconductors. S.G. is solely responsible for the section about the chirality rule in the supplement. S.G. wrote the first version of the manuscript. J.G.R, A.C.B and J.W. contributed to the measurements and data analysis.

S.G., J.G.R., A.C.B., J.W., M.T., and I.K.S. interpreted the results and wrote the manuscript.

# Supplementary Materials: Search for New Superconductors: An Electro-Magnetic Phase Transition in an Iron Meteorite Inclusion at 117 K

S. Guénon[1,2*], J. G. Ramírez[1,3], Ali C. Basaran[1], J. Wampler[1], M. Thiemens[4], and Ivan K. Schuller[1]

**Affiliations:**

[1] Department of Physics and Center for Advanced Nanoscience, University of California San Diego, La Jolla CA 92093, USA.

[2] CQ Center for Collective Quantum Phenomena and their Applications in LISA+, Physikalisches Institut, Eberhard Karls Universität Tübingen, Auf der Morgenstelle 14, D-72076, Tübingen, Germany.

[3] Department of Physics, Universidad de los Andes, Bogotá 111711, Colombia

[4] Department of Chemistry and Biochemistry, University of California San Diego, La Jolla CA 92093, USA.

**Experimental Details**

A detailed description of the Magnetic Field Modulated Microwave Spectroscopy (MFMMS) technique can be found elsewhere [28,29]. Here, we want to point out two important experimental details. First, there is a thermal hysteresis between the cooling and heating curves at ramp rates of 2K/min due to thermal backlash of the continuous flow cryostat. This hysteresis is considerably reduced at slow ramp rates.

Second, when we reinstalled a sample tube, certain characteristics like absolute values, or the ratio between the peak height at 110 K and the background had changed. However, important features like the peak position or the overall shape of the background were reproducible. This could be due the fact that MFMMS is a surface sensitive technique. When a sample tube is removed and remounted in the system the arrangement of the particles in the quartz tube could change and consequently the response in MFMMS. There was one case were the peak at a 15 Oe DC offset field had disappeared completely after the sample was magnetized with 9000 Oe. In this case the peak was only visible in MFMMS-spectra with a DC offset field of 50 Oe and higher. Probably, the remnant magnetization of the ferromagnetic matrix was reducing the response in MFMMS.

## How to distinguish between a diamagnetic and paramagnetic response with hysteretic magnetic field scans

In this section we will discuss that the distinguishing characteristics in the spin-glass behavior of superconducting and ferromagnetic materials reported in Sastry et al. [34] is the manifestation of a more general rule.

In the low field microwave absorption study on ferromagnetic $Gd_2CuO_4$, Sastry et al. have found that the contrasting feature between the hysteretic magnetic field sweeps of cuprate superconductors and the ferromagnetic spin-glass is the relative position of the signal during field increasing and decreasing cycles. An equivalent statement would be that the chirality is different between the magnetic field sweep loops of the superconductors and of the ferromagnet.

Indeed, based on our experience with the MFMMS-technique and extensive literature studies, we can confirm following rule that distinguishes between a diamagnetic response of superconductors and the paramagnetic response of ferromagnetic materials: For a diamagnetic response the chirality of hysteretic field scan loops (FSLs) is clockwise, while it is counter-clockwise for a paramagnetic response.

The chirality rule can only be applied, if the there is no background and the response is positive, i.e. the phase transition under investigation is causing a peak and not dip in the temperature scan. If the MFMMS-spectrum has a background, it needs to be subtracted, and if the signal is negative, the sign of the response needs to be inverted, before the chirality rule can be applied.

In the following we will explain the physics responsible for the chirality rule.

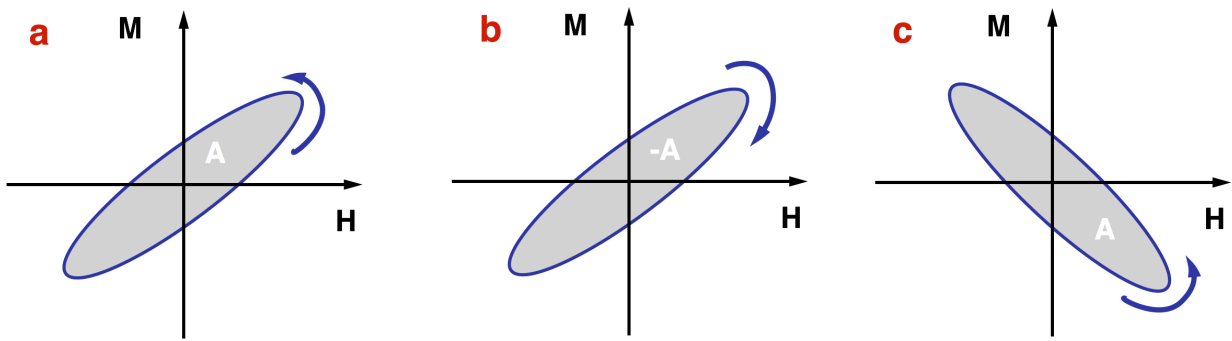

FIG. S1. Chirality rule. (a) Idealized illustration of the hysteretic magnetization loop corresponding to a ferromagnetic material. (b) Illustration of the magnetization loop of an unknown material. (c) Idealized illustration of the magnetization loop corresponding to a superconductor.

We first discuss a simplified case: instead of hysteretic MFMMS-FSLs, we consider hysteretic magnetization loops as they might be acquired with a magnetometer. In the Fig. S1(a), an idealized illustration of a hysteretic magnetization curve corresponding to a ferromagnetic material is shown. The energy dissipated in the material during one field sweep cycle is proportional to the area A enclosed by the magnetization loop. We now consider an "unknown" material with an idealized magnetization loop almost identical to the one from the ferromagnet with the difference that the chirality is opposite [Fig. S1(b)]. If such a material would exist, it would imply that it produces energy during a magnetization cycle. This would be unphysical, and therefore such a material cannot exist. However, if we assume, that for some technical reasons, we have lost the information about the correct sign of the magnetization than we could learn from this magnetization loop, that we have to invert the sign of magnetization to avoid unphysical results, and consequently the correct magnetization loop should look like the one depicted in Fig. S1(c). This is the idealized magnetization loop of a material with strong diamagnetism, i.e. a superconductor. To conclude this paragraph: it is possible to recover the correct sign of the magnetization from the magnetization loop chirality, and this way a diamagnetic response can be distinguished from a paramagnetic response.

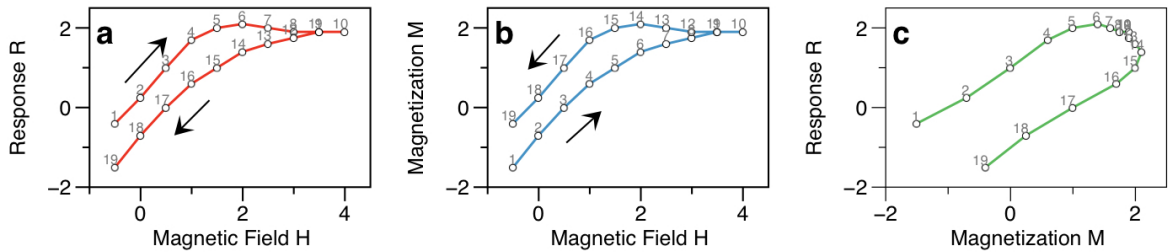

FIG. S2. Chirality flip. (a) an idealized MFMMS field scan loop as it was reported in the main article. (b) Magnetization curve identical to (a), but with opposite chirality. (c) Response vs. Magnetization derived from (a) and (b).

We now return to the MFMMS-technique: in MFMMS the response R is proportional to the derivative of the absorbed microwave power with respect to the magnetic field: $R \propto dP/dH$. This means that the MFMMS-response R of a material depends on an intrinsic material property that can be influenced by an external magnetic field. In most cases it is safe to assume that the response R depends on the magnetization of the material under investigation. However, because different materials have different response mechanisms, it is not possible to directly derive the sign of the magnetization from the sign of the MFMMS-signal, and therefore we cannot distinguish a diamagnetic response from a paramagnetic response. For instance, the sign of the response caused by ferromagnetic GaMnAs is equal the sign of a superconductor response [28].
However, in the preceding section we have learned, how to recover the sign from hysteretic magnetization loops, and, under the assumption that the MFMMS-response depends on the magnetization and that the response mechanism preserves the chirality, this procedure can be applied to a MFMMS-FSL, too.

To address the question, under which conditions a response function preserves the chirality, we discuss a counter-example: Figure S2(a) shows an idealized FSL, which is similar to the ones of the main paper. Assume that in this case the material has a MFMMS-response mechanism that only changes the chirality, i.e. the corresponding magnetization curve has the identical shape but the chirality is reversed [Fig. 2(b)]. The response function vs magnetization R(M) can be easily extracted by plotting the response vs. the magnetization for every individual point [Fig. 2(c)]. We learn that, if the chirality is not preserved in MFMMS, the response function itself has to be hysteretic.

For most materials the MFMMS-response seems to depend on the magnetization in a well-defined, non-hysteretic way, and the chirality rule applies. However, there could be exceptions. For instance, in nano-magnetic systems the complex interplay between the bulk-magnetization and the surface magnetization can cause a response in the surface sensitive MFMMS-technique, which cannot be explained in the framework of a response mechanism that only depends on the bulk magnetization. So far, we are aware of only one anomalous case, where the chirality rule does not apply. The exception is a microwave study on fine cobalt ferrite particles [38]. The up-sweep curve of the FSL in Fig. 2 in this article resembles the up-sweep curves of the spin-glass studies [33-35], however the down sweep curve is different: it has the shape of an up-sweep curve that is mirrored on the x-axis. Further, it is important to note, that the up-sweep curve maximum is at higher field values than the down-sweep minimum. This behavior is typical of a ferromagnetic system.

In Fig. 3 of this article the chirality of the FSLs was not indicated. The authors did not comment on the details of the FSLs and we cannot explain them with the limited information provided in this article. Consequently, the reason for the deviation from the chirality rule in this specific case is elusive, but it could be due to the interplay between bulk and surface magnetization mentioned above.

Although there is an exception and the chirality rule is not universal, i.e. it is not hundred percent save to assume that it applies for all known and unknown materials, the chirality of the BRFSLs in Fig. 3 of the supplemented article is very strong evidence that the reported phase transition is caused by a superconductor, because it would be very unlikely that a material would violate the chirality rule while retaining all the other reported [33-35] field sweep characteristics of a vortex glass state.

**Materials:**

Iron meteorites, containing predominantly Fe and Ni, are thought to be crystallized cores of protoplanetary bodies [11-15]. They formed from melting chondritic material and separation of Fe and Ni as the melt segregated from silicate. In the case of Group IVB irons, based upon measurements of P, V, Cr, Fe, Co, Ni, Cu, Ga, Ge, As, Mo, Ru, Rh, Pd, W, Re, Os Ir, Pt and Au, models of potential core formation mechanisms reveal that no single temperature evolution explains the volatility trend (e.g. Campbell et al. [11]). Besides the original segregation of Fe and Ni in the pre solar nebula, there is evidence for secondary oxidative processes required during planetary differentiation to eliminate elements such as Cr, Ga, and W, as observed in group IVB meteorites such as Tlacotepec. The group IVB iron meteorites, measured in the present work, are ataxites with Widmanstädten patterns of small kamacite platelets indicating high bulk Ni content, typically 15-18%, consistent with a secondary oxidizing environment [13]. Measurements of $^{107}Ag/^{109}Ag$ ratios in iron meteorites, including Tlacotepec, have shown enrichment with respect to terrestrial ratios. The excess is attributed to *in situ* decay of extinct $^{107}Pd$ ($\tau_{1/2}$ = 6.5 x $10^6$ yr) from a late nucleosynthetic event near the formation time of the solar system [17,18]. The correlation of $^{107}Ag/^{109}Ag$ with $^{107}Pd$ requires a short time period (<$10^7$ years) to allow in situ formation of $^{107}Pd$ in the iron core of a small differentiated planetary object [19]. Sulfur isotopic anomalies are found in troilite, schreibersite, and FeNi alloys. Excess $^{33}S$ and $^{36}S$ [39,40] has been observed in iron meteorites and their composition attributed to high energy cosmic ratio spallation of Fe-Ni of the iron meteorite parent at billion year time scale requiring a long exposure to cosmic rays. In the case of Tlacotepec, a cosmic ray exposure of the Fe-Ni core shielded within a planet for a time scale of approximately 800 million years [39,40]. No evidence of excess $^{33}S,^{36}S$ is observed in the Troilite [40].

Troilite is the most common inclusion in iron meteorites with formula of $Fe_{1-x}S$ with x varying between 0 and 0.2. In Tlacotepec, Troilite may occur as an occasional 1-2 cm nodule, and as millimeter –sized particles of box or plate shape with a frequency of approximately 15 $cm^{-1}$ [16]. Troilite is normally nonmagnetic but if melted and cooled becomes magnetic. One of the most extensive studies of the mineralogy of iron meteorites was for group III AB irons [20]. The mineralogy is reported as being extraordinarily diverse and includes, chromite, sphalerite, phosphate bearing sulfides, silica and native copper. Four different phosphate bearing minerals are observed, 6 sulfidic phases, and other rare minerals such as CrN. Sulfide nodules are suggested as originating as droplets of an immiscible sulfur rich liquid that continuously segregated from the parent melt during crystallization of the melt, consistent with recent models. Though this extensive work is for group IIIAB, it is reasonable to assume that these minerals are present in IVB mineral but an equivalent detailed analysis is not available. The trace species associated with the iron meteorites vary (Fe/Ni, and Ga, Ge), it is not unreasonable to qualitatively assume the two to be mineralogical similar.

The thermal history of group IVB meteorites is typical of ataxite meteorites. They possess microscopic Widmanstädten patters consisting of kamasite platelets embedded in platelets of plessite [13]. The Tlacotepec meteorite consists of 15.9 % Ni, and 0.05 % Pt. The inferred kamasite nucleation temperature is 4745 degree centigrade, with a cooling rate of 500 degrees per million years [13]. The meteorite resided as a core within a larger asteroidal body at a pressure of <1 GPa and was released following a glancing impact from a larger asteroidal body. The IVB parent body attained a high internal temperature of 1760 K or greater which lead to a homogenization and rapid formational age. Chen and Wasserburg (1990) on using $^{107}Ag/^{108}Pd$ have shown that this meteorite formed some 34 million years earlier than various iron meteorites [41]. Kaiser and Wasserburg (1983) have measured the $^{107}$Ag/$^{109}$Ag ratios in Troilite and the surrounding host metals [42]. Differences between the phases for Grant and Santa Clara suggest redistribution between the phases during cooling. Carlson and Hauri (2001) have applied multi collector plasma ionization mass spectrometry to determine $^{107}$Ag/$^{109}$Ag at higher precision [43] and shown that the difference between sulfide inclusions and the iron host clearly are statistically present in Canyon Diablo and is nearly identical to that in Gibeon observed by Chen and Wasserburg (1990) and Kaiser and Wasserburg (1983).

**Comparison between MFMMS, SQUID and AC-susceptibility:**

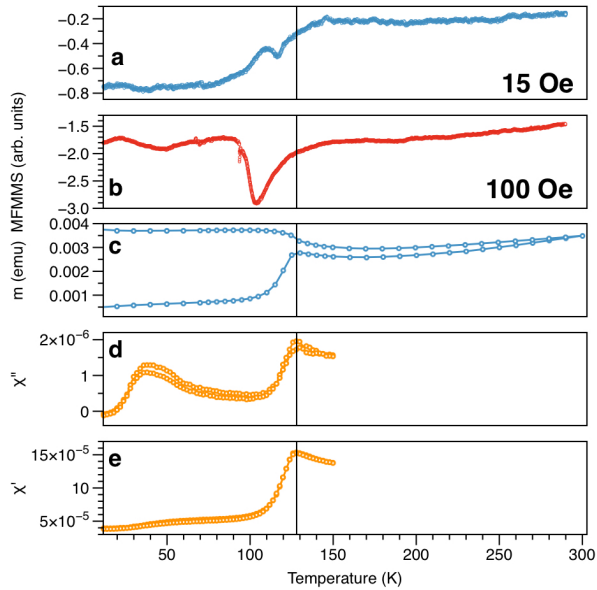

FIG. S3. Comparison between MFMMS (a,b), SQUID (c) and AC-susceptibility (d,e) spectra of Tlacotepec sample (Tla-3). (a) MFMMS cooling, magnetic DC-field 15 Oe. (b) MFMMS heating, magnetic DC-field 100 Oe. c, SQUID: zero field, non-zero field curves (100 Oe). (d, e) AC-susceptibility spectra.

MFMMS, SQUID and AC-susceptibility spectra were acquired from a Tlacotepec sample [Fig. S3]. The MFMMS with a magnetic DC-field of 15 Oe [Fig. S3a] shows the typical negative background step with the onset at 140 K and a peak with the onset at 117 K. Note that this spectrum is a cooling curve, while the spectrum for a 100 Oe DC-field is a heating curve [Fig. S3(b)]. In this case the onset of the background step is at 130 K and the peak onset is at 102 K. The zero field curves and field cooled curves of the SQUID are typical for a ferromagnetic phase transition with a $T_c$ of 130 K. The same is true for the AC-susceptibility measurements. In the $\chi''$ curve is a bump at 50 K indicating a significant change in magnetic losses in a temperature range where the other data do not indicate a phase transition.

These results raise the question whether the ferromagnetic transition observed with the SQUID and AC-susceptibility measurements is related to the peak in MFMMS. Both SQUID and AC-susceptometer are commercial and calibrated systems, and the scan time for one temperature sweep was several hours. Hence, the systems had enough time to thermalize before a data value was acquired. On the contrary, the scan time in the MFMMS was only one hour. When the sweep time is reduced like for some measurements in the main paper the peak onset position is at 117 K. Therefore, we consider it as unlikely that the peak is related to the ferromagnetic transition at 130 K. It is more likely that the SQUID and the AC-suceptometer were not sensitive enough to detect phase transition related to the MFMMS peak.

**Energy dispersive X-ray spectroscopy**

We used a FEI-XL30 scanning electron microscope with an EDX-detector to gather first information about the materials composition. We decided to investigate a sample with electrical contacts, because this way charging effects due to the electron beam are avoided. The area between the silver paint contacts was chosen for imaging [see Fig. S4]. Besides Ag from the silver paint, only Fe and S could be detected. The materials distribution maps show that the stoichiometry of Fe and S varies slightly in the samples. These results confirm that the materials under investigation are predominately iron-sulfides like Troilite. However, superconductivity is very sensitive to the exact stoichiometry and trace elements can significantly influence this material property. Consequently, for a targeted investigation of the chemical compositions more sophisticated techniques are needed.

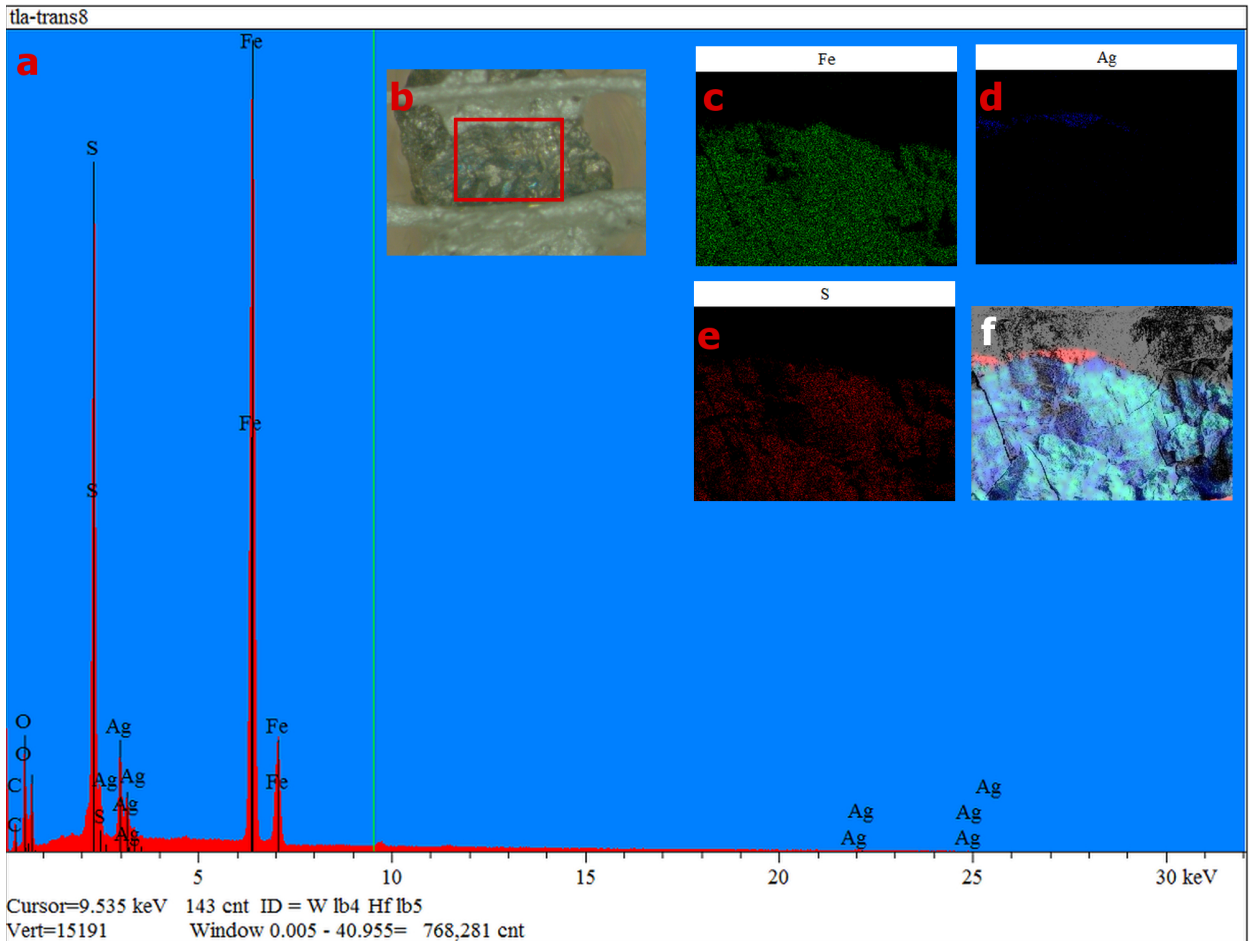

FIG. S4. Tla-6, EDX-analysis. (a) spectrum. Insets: (b) optical micrograph. Field of view is indicated by the red rectangle. The width of the red rectangle is 0.75 mm. (c-e) iron, sulfur and silver distribution, respectively. (f) overlay image.